\documentclass[12pt,a4paper,twoside]{paper}
\usepackage{appendix}
\usepackage{chngcntr}
\usepackage{etoolbox}
\usepackage{lipsum}
\usepackage{amsmath,bbm,amssymb,amsxtra}
\usepackage{enumerate}
\allowdisplaybreaks
\usepackage[active]{srcltx}
\usepackage{epsfig}
\usepackage{ae}
\numberwithin{equation}{section}

\newtheorem{lemma}{Lemma}[section]
\newtheorem{proposition}[lemma]{Proposition}

\newtheorem{theorem}[lemma]{Theorem}

\newtheorem{definition}[lemma]{Definition}

\newcommand{\prop}[1]{\begin{proposition}\label{#1}
\sl }
\newcommand{\eprop}{\end{proposition}}
\newcommand{\thm}[1]{\begin{theorem}\label{#1}
\Ä }
\newcommand{\ethm}{\end{theorem}}
\newcommand{\lem}[1]{\begin{lemma}\label{#1}
\sl }
\newcommand{\elem}{\end{lemma}}

\newcommand{\defin}[1]{\begin{definition}\label{#1}
\sl }
\newcommand{\edefin}{\end{definition}}

\newcommand{\beqno}{\begin{eqnarray*}}
\newcommand{\eeqno}{\end{eqnarray*}}
\def\eeq {\end {eqnarray}}
\newcommand{\beq}{\begin {eqnarray}}

\def\qq{ \begin{eqnarray} }
\def\qqq{ \end{eqnarray} }
\def\rr{ \begin{equation} }
\def\rrr{ \end{equation} }

\newcommand{\C}{{{\mathbb{C}}}}

\def\qq{ \begin{eqnarray} }
\def\qqq{ \end{eqnarray} }

\newcommand{\BbbR}{\mathbb{R}}

\newcommand{\be}{\begin{equation}}
\newcommand{\ee}{\end{equation}}

\newcommand{\ba}{\begin{eqnarray}}
\newcommand{\ea}{\end{eqnarray}}

\newcommand{\bea}{\begin{eqnarray*}}
\newcommand{\eea}{\end{eqnarray*}}

\newcommand{\bee}{\begin{enumerate}}
\newcommand{\ene}{\end{enumerate}}

\usepackage{epsfig}

\counterwithin*{equation}{section} 
\counterwithin*{equation}{subsection} 

\begin{document}

\pagestyle{empty}

     \vspace*{10mm}

     \begin{center}
 \textsl{\LARGE    History of Quantum Mechanics or  the Comedy of Errors}\footnote{Talk given on  November 11, 2014,
 at the conference: La m\'ecanique \`a la lumi\`ere de son histoire, de la Modernit\'é \`a l'\'epoque contemporaine (XVIIe-XXe s.), given in honor of Prof. Patricia Radelet -De Grave.}

 \vspace*{10mm}
 { \LARGE Jean BRICMONT

IRMP,
UniversitŽ catholique de Louvain, 
chemin du Cyclotron 2,

1348 Louvain-la-Neuve,
Belgique}
\end{center}
\begin{abstract}
The goal of this paper is to explain how the views of Albert Einstein, John Bell and others, about nonlocality and the conceptual issues raised by quantum mechanics, have been rather systematically misunderstood by the majority of physicists.

\end{abstract}

\section{Introduction}\label{sec1}

The history of quantum mechanics, as told in general to students, is like a third rate  American movie: there are the good guys and the bad guys, and the good guys won.

The good guys are those associated with the ``Copenhagen" school, Bohr, Heisenberg, Pauli, Jordan, Born, von Neumann among others. The bad guys are their critics, mostly Einstein and Schrödinger and, sometimes de Broglie. The bad guys, so the story goes, were unwilling to accept the radical novelty of quantum mechanics, either its intrinsic indeterminism or  the essential role of the observer in the laws of physics that quantum mechanics implies. They were hinged to a classical world-view, because of their philosophical prejudices.

Einstein invented various {\it gedanken experiments} in order to show that the Heisenberg uncertainty relations could be violated (those relations put limits on what can be known about physical systems), but Bohr successfully answered those objections. After the war, in 1952, David Bohm rediscovered an old idea of de Broglie and tried to develop an alternative to quantum mechanics that would restaure determinism, but his theory got very little attention among physicists, mostly because, by then, the successes of quantum mechanics had been so spectacular that hardly anybody had still doubts about its correctness.

Bohm's theory was introducing ``hidden variables", meaning some unobservable quantities (hence, ``metaphysical") that are not part of the standard quantum mechanical  description.\footnote{Throughout this paper, we will use the expression ``hidden variables" to mean any variables that are used to describe the state of a quantum system and that are not included in the usual quantum description, via a wave function or a quantum state (the latter being simply a wave function ``times" some internal states such as spin states, but we will not stress that distinction here and we will speak only of wave functions).} Einstein had also favored the introduction of such variables in order to ``save determinism". The first objection to this idea is obvious: why bother with unobservable entities in order to satisfy a philosophical prejudice?

However, the definite blow against hidden variables was given in 1964 by John Bell who showed that, merely imagining that such variables exist leads to predictions that are in contradiction with those of quantum mechanics. Moreover, those specific predictions were later tested in laboratories and, of course, the observations came definitely on the side of quantum mechanics and against hidden variables. Case closed!

The goal of this paper is to show that all of the above is essentially false.\footnote{The material for this paper is in part taken from \cite{Bri, Bri1}, and we refer to those texts for more details. In particular, we will not discuss much Bohm's theory here, but only note that it is not true that this theory has been refuted Bell; see  \cite{Bo1, DT, DGZ, Go-St, Bri} for an
exposition of that theory.} And in order to show that, we will have to do almost no physics, just reading  what Einstein and Bell really said (as Patricia always told me: ``go read the sources!"). One may not agree with them, one may adhere to the standard views on quantum mechanics (whatever they are, which nowadays is not so simple to say), but one should, it seems to me, at least try to understand {\it what Einstein and Bell  said before rejecting their ideas}!

We shall first explain what really worried Einstein about quantum mechanics, namely its nonlocal aspect.\footnote{At least that was one of his worries. He also worried about the centrality of the ``observer" in quantum mechanics, but we will not discuss that problem here (see \cite{Bri}). Let us nevertheless mention that this worry is illustrated by the following remark by Abraham Pais about his conversations with Einstein \cite[p. 907]{Pais}: ``We often discussed his notions on objective reality. I recall that during one walk Einstein suddenly stopped, turned to me and asked whether I really believed that the moon exists only when I look at it.'' ``Is the moon there when nobody looks?" has also become a famous Einstein quote (frequently misunderstood, since Einstein had no doubt that the answer to that question was ``yes", while quantum mechanics is often interpreted as implying that we don't know anything about the world before we ``look").} Next, we will discuss what Einstein (together with Podolsky and Rosen) and Bell said about that issue of nonlocality. Then we will review how this was misunderstood by most physicists.

\section{Einstein's Real Worries}\label{sec1a}

The most often quoted phrase of Einstein is probably ``God does not play dice".\footnote{The complete quote comes from a letter to Max Born in 1926 \cite[p.~91]{Bor}: ``Quantum mechanics is very worthy of regard. But an inner voice tells me that this is not yet the right track. The theory yields much, but it hardly brings us closer to the Old One's secrets. I, in any case, am convinced that He does not play dice." Of course, as Einstein emphasized several times, his ``God" had nothing to do with the personal gods of the ``revealed" religions.}  But this is misleading. It is true that Einstein did express concerns about determinism, but that determinism was Einstein's main concern was   strongly denied even  by Pauli, who was, in general, on the Copenhagen side of the arguments (if one wants to use this somewhat reductionist terminology) and who wrote in 1954 to Max Born:

\begin{quote}

Einstein does not consider the concept
of ``determinism" to be as fundamental as it is frequently held to
be (as he told me emphatically many times) [\dots] he disputes that he uses
as a criterion for the admissibility of a theory the question: ``Is it
rigorously deterministic?"

\begin{flushright} Wolfgang Pauli \cite[p.~221]{Bor} \end{flushright}
\end{quote}

If determinism was not Einstein's main concern, what was? A clue to the answer is given in a letter, written in 1942, which also speaks of God not playing dice:

\begin{quote}

It seems hard to sneak a look at
God's
cards. But that he plays dice and uses ``telepathic'' methods (as the present quantum theory requires of him)
is something that I cannot believe for a single moment.

\begin{flushright} Albert Einstein \cite{EinstL} quoted in \cite[p.~68]{DH} \end{flushright}

\end{quote}

What Einstein called ``telepathic" here is related to what nowadays is called nonlocality and this is the  aspect of quantum mechanics that bothered Einstein. He and Schrödinger  saw clearly that aspect when everybody else had their heads in the sand, so to speak.

To understand nonlocality, let us think for a moment about the concept of wave function and what it means to say (as is usually done) that it provides a {\it complete description} of quantum systems. The wave function, let's say for one particle, is a (complex valued) function defined on $\BbbR^3$: $\Psi(x)\in \C$, $x \in \BbbR^3$. Its meaning, in quantum mechanics textbooks is that the square of its absolute value, $|\Psi(x)|^2$, gives the probability density of {\it finding} the particle at a given point, if one ``measures" the particle's position.  A naive interpretation of the word measurement would lead us to think that, if the particle is found somewhere, it is because {\it it was there} in the first place! But that is not what quantum mechanics says: indeed, ``measurements" might affect the object being measured and not simply reveal any of its properties, so that the word measurement is used in quantum mechanics in a rather awkward way.

But what worried Einstein is that, after the measurement of the position, the wave function changes and collapses to a wave function concentrated at or around the point where the particle is found. But that means that the value of the wave function suddenly jumps to zero everywhere, except where the particle is found and that effect is supposed to be instantaneous. Since the wave function can be, before the measurement, different from zero far away from the point of detection, it means that the collapse of the wave function is a sort of nonlocal action, or an action at a distance, and this is what Einstein could not accept (in part, because it contradicted the theory of relativity, as he and most people understood it).\footnote{That is because the relativity of simultaneity, which is part of the theory of relativity, implies that, if causal effects can be instantaneous in some reference frames, then they must act backward in time in some other reference frames. This, to put it mildly, contradicts our intuitive notion of causality. This problem is discussed in detail in \cite{Ma}.} That action at a distance is what ``telepathy" means in the above quote;\footnote{To be precise, Einstein was probably referring there to the stronger sort of nonlocality implied by the EPR argument discussed in Sect.~\ref{sec5}.} Einstein also called such actions at a distance ``spooky" (\cite[p.~158]{Bor}).

\begin{figure}[t]
\centering
\includegraphics[width=.65\textwidth,angle=-1]{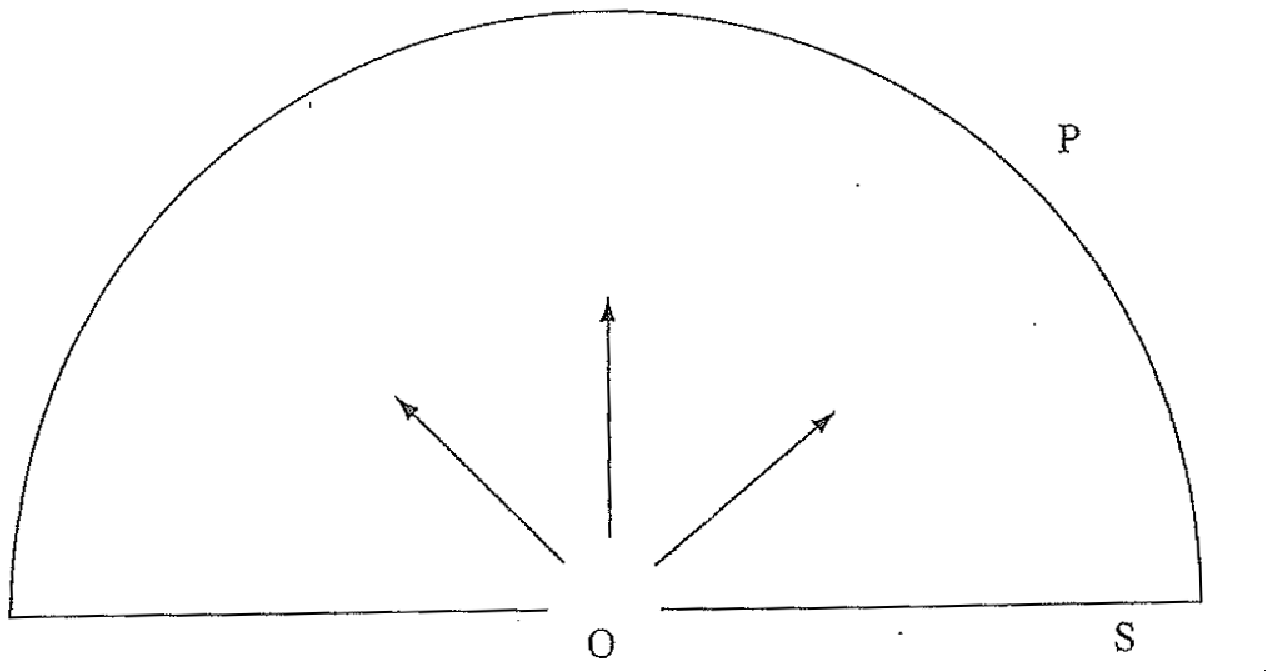}
\caption[]{Einstein's objection at the 1927 Solvay Conference. Reproduced with the permission of the Solvay Institutes and of Cambridge University Press. See \cite[p.~440]{BV}, or \cite[p.~254]{Sol} for the ``original" (published in French translation at the time of the Solvay Conference). In fact Einstein raised a similar issue as early as 1909, at a meeting in Salzburg \cite[p.~198]{BV}}\label{7fig1}
\end{figure}

Already at the 1927 Solvay Conference, Einstein expressed very clearly this worry. He considered a particle going through a hole, as shown in Fig.~\ref{7fig1}. In the situation described in the picture, the wave function spreads itself over the half circle, but one always detects the particle at a given point, on the hemispherical photographic film denoted by $P$ in Fig.~\ref{7fig1}. If the particle is {\it not\/} localized anywhere before its detection (think of it as a sort of ``cloud", as spread out as the  wave function), then it must condense itself at a  point, in a nonlocal fashion, since the part of the particle that is far away from the detection point must ``jump" there instantaneously. If, on the other hand, the particle  {\it is\/} localized somewhere before its detection, then quantum mechanics  {\it is\/} an  incomplete description of reality, since quantum mechanics does not include the position of the particle in its description.

Einstein contrasted two conceptions of  what the  wave function could mean:
 
\begin{quote}

Conception I. The de Broglie--Schr\"odinger waves do not correspond to a single electron, but to a cloud of electrons extended in space. The theory gives no information about individual processes, but only about the ensemble of an infinity of elementary processes.

Conception II. The theory claims to be a complete theory of individual processes. Each particle directed towards the screen, as far as can be determined by its position and speed, is described by a packet of de Broglie--Schr\"odinger waves of short wavelength and small angular width. This wave packet is diffracted and, after diffraction, partly reaches the screen P in a state of resolution.\footnote{That last expression is not very clear, but the original German text has been lost and the French translation \cite{Sol}, from which this is translated into English, is not clear either. (Note by J.B.)}



\begin{flushright} Albert Einstein \cite[p.~441]{BV} \end{flushright}
\end{quote}
The distinction made here was repeated by Einstein all his life. Either one adopts a statistical view (conception I) and a more complete theory can be envisioned (for ``individual processes"), or one declares the quantum description ``complete", but this means that the particle is spread out in space before being detected on the screen.

After explaining the merits of conception II, Einstein raises the following objection:

\begin{quote}

But the interpretation, according to which $|\psi|^2$ expresses the probability that {\it this} particle is found at a given point, assumes an entirely peculiar mechanism of action at a distance, which prevents the wave continuously distributed in space from producing an action in {\it two} places on the screen.

\begin{flushright} Albert Einstein \cite[p.~441]{BV} \end{flushright}
\end{quote}
Indeed, if the particle is spread out in space before being detected (which is what the expression ``complete description" {\it means}), then the fact that it is always detected at a given point implies that it condenses itself on that point and that its presence vanishes elsewhere. Thus something nonlocal must be taking place. Einstein adds:

\begin{quote}

In my opinion, one can remove this objection [action at a distance] only in the following way, that one does not describe the process solely by the Schr\"odinger wave, but that at the same time one localises the particle during the propagation. I think that Mr de Broglie is right to search in this direction.\footnote{Einstein refers here to de Broglie's guiding wave theory, which was also presented  at the 1927 Solvay Conference, see  \cite{BV} or \cite[chap. 8]{Bri} for a further discussion of that Conference. The de Broglie's guiding wave theory was later developed by Bohm.} If one works solely with the Schr\"odinger waves, interpretation II of $|\psi|^2$ implies to my mind a contradiction with the postulate of relativity.

\begin{flushright} Albert Einstein \cite[p.~441]{BV} \end{flushright}
\end{quote}
Thus, the essence of Einstein's objection to the orthodox view of quantum mechanics (including the conflict between nonlocality and relativity) was already expressed in 1927! 

It is interesting to read Bohr's response:

\begin{quote}

I feel myself in a very difficult position because I don't understand what precisely is the point which Einstein wants to [make]. No doubt it is my fault.

\begin{flushright} Niels Bohr \cite[p.~442]{BV} \end{flushright}
\end{quote}
Einstein's objections became sharper as time went on, specially with the 1935 Einstein, Podolsky, and Rosen  (or EPR) paper  \cite{EPR}. 

Before turning to the EPR argument, we will start by a little known, but very simple, thought experiment, known as Einstein's boxes. This example will allow us to raise and explain the issue of locality. Then we will define precisely what we mean by nonlocality and give a simple derivation of Bell's argument (due to \cite{DGTZ}), {\it combined} with the EPR argument. That is the simplest and clearest way to arrive at the main conclusion, namely that the world is nonlocal in a sense made explicit in Sect.~\ref{sec3}. 

\section{Einstein's Boxes}\label{sec2}

\begin{figure}[t]
\centering
\includegraphics[width=.7\textwidth]{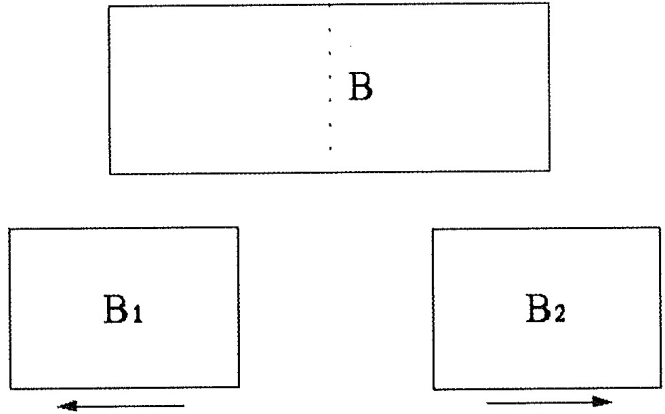}
\caption[]{Einstein's boxes. Reproduced with permission from T. Norsen: Einstein's boxes, {\it American Journal of Physics} {\bf 73}, 164--176 (2005). Copyright  2005 American Association of Physics Teachers}\label{4fig1}
\end{figure}

Consider the following thought experiment.\footnote{We base ourselves in this section on \cite{TN}, where the description of the experiment is due to de Broglie \cite{dB1, dB}. Einstein's original idea was expressed in a letter to Schr\"odinger, written on 19 June 1935, soon after the EPR paper was published \cite[p. 35]{Fi}. However, Einstein formulated his ``boxes" argument in terms of macroscopic objects (small balls) and then gave a slightly different and genuinely quantum mechanical example, but illustrating the same point as the one made here. Figure~\ref{4fig1} is taken from \cite{TN}.} There is a single particle in a box $B$ (see Fig.~\ref{4fig1}), and its wave function  is $ |B\rangle$, meaning that its wave function is distributed\footnote{The precise distribution does not matter, provided it is spread over $B_1$ and $B_2$ defined below; it could be distributed according to the square of the ground state wave function of a particle in the box $B$.} over $B$. One cuts the box into two half-boxes, $B_1$ and $B_2$, and the two half-boxes are then separated and sent as far apart as one wants.
 
According to ordinary quantum mechanics, the wave function becomes
\[
\frac{1}{\sqrt{2}}\big(| B_1\rangle + |B_2\rangle\big)\;,
\]
where the wave function $ |B_i \rangle$ means that the particle ``is" in box $B_i$, $i=1,2$ (and again, each wave function $| B_i\rangle$ is distributed in the corresponding box).\footnote{This is of course a thought experiment. We assume that we can cut the box in two without affecting the particle. But similar experiments have been made with photons and were already suggested by Heisenberg in 1929, see \cite[p. 39]{Heis0}.} Here, we put scare quotes around the verb ``is" because  of the ambiguity inherent in the meaning of the wave function, reflected in the two notions discussed by Einstein at the 1927 Solvay Conference:  if the wave function reflects our knowledge of the system, then the particle {\it is} in one of the boxes $B_i$, without quotation marks. But if one thinks of the wave function as being physical and of the position of the particle as being created or realized only when someone measures it, then the quotation marks are necessary and ``is" means: ``would be found in box $B_i$ after a  measurement of its position". 

Now, if one opens one of the boxes (say $B_1$) and one does {\it not\/} find the particle in it, one {\it knows\/} that it is in $B_2$. Therefore, the wave function ``collapses'' instantaneously: it becomes  $|B_2\rangle$ (and if one opens box $B_2$, one will find the particle in it!).

Since $B_1$ and $B_2$ are spatially separated (and as far apart as we wish), if we reject the notion of action at a distance, then it follows that acting on $B_1$, namely opening that box, cannot have any physical effect whatsoever on $B_2$. However, if opening box $B_1$ leads to the collapse of the wave function into one where the particle is necessarily in $B_2$, it must be that the particle was in $B_2$ all along. That is, of course, the common sense view and also the one that we would reach if the particle was replaced by any large enough object.

But in the situation of the particle in the box, if we reject action at a distance, then we must admit that quantum mechanics is not complete, in the sense that Einstein gave to that word: there exist other variables (called ``hidden variables") than the wave function that describe the system, since the wave function does not tell us in which box the particle is  and we just showed, assuming no action at a distance, that    
the particle {\it is\/} in one of the two boxes, before one opens either of them.

In any case, with this argument, Einstein had already proven the following dilemma: either there exists some action at a distance in Nature (opening box $B_1$ changes the physical situation in $B_2$) or quantum mechanics is incomplete. Since action at a distance was anathema to him (and probably to everybody else at that time), he thought that he had shown that quantum mechanics is incomplete.

There are many examples at a macroscopic level that would raise a similar dilemma and where one would side with Einstein in making assumptions, even very unnatural ones, that would preserve locality. Suppose that two people are located far apart, and each tosses coins with results that are always either heads or tails, randomly, but always the same for both throwers. Or suppose that in two casinos, far away from each other, the roulette always ends up on the red or black color, randomly, but always the same in both casinos. Or imagine twins far apart that behave exactly in the same fashion.\footnote{This last example is given by Bell in his interview with Jeremy Bernstein \cite[p. 63]{Be}.} In all these examples (and in many others that are easy to imagine), one would naturally assume (even if it sounded very surprising) that the two coin throwers or the casino owners were able to manipulate their apparently random results and coordinate them in advance or, for the twins, one would appeal to a strong form of genetic determinism. Who would suppose that one coin tosser immediately affects the result of the other one, far away, or that the spinning of the ball in one casino affects the motion of the other ball, or that the action of one twin affects the behavior of the other twin? In all these cases, one would assume a locality hypothesis; denying it would sound even more surprising than whatever one would have to assume to explain those odd correlations.

But one thing should be a truism, namely that those correlations pose a dilemma: the results are either coordinated in advance or there exists a form of action at a distance. Therefore, Einstein's conclusion in the case of the boxes (incompleteness of quantum mechanics), which is similar to the assumptions we made here about coin throwers, casinos, and twins, was actually very natural.

As an aside, let us mention that the example of the boxes also raises a serious question about the transition from quantum to classical physics. Indeed, if the quantum particle is replaced by a ``classical'' one, meaning a large enough object, nobody denies that the particle {\it is\/} in one of the boxes before one opens either box. But where is the dividing line between the quantum realm and the classical one? The transition from quantum to classical physics is usually thought of as some kind of limit, like considering large masses or large energies (compared to the ones on the atomic scale); but a limit is something that one gets closer and closer to when a parameter varies. Here, we are supposed to go from the statement ``the particle is in neither of the boxes" to ``the particle is in one of them, but we do not know which one". This is an ontological jump and not the sort of continuous change that can be expressed by the notion of limit.

Let us now define more precisely what we mean by nonlocality.

\section{What Is Nonlocality?}\label{sec3}

Let us consider what kind of nonlocality or actions at a distance would be necessary, in the example of the boxes, in order to deny Einstein's conclusion about the incompleteness of quantum mechanics. So assume that the particle is in neither box, before one opens one of them (which is what quantum mechanics being complete means). Then opening one box, say $B_1$, creates the particle, either in $B_1$ or in $B_2$, and consider the latter possibility here. This would nonlocally create a particle in the unopened box $B_2$, and that action at a distance would have the following properties:
\begin{itemize}
\item[1.]  {\it That action is instantaneous\/}:\footnote{Of course, instantaneity is not a
relativistic notion, so let us say, instantaneous in the reference frame where both boxes are at rest. We will not discuss relativity here, see 
\cite[pp.~136--137]{Ma} for a discussion of the conflict between relativity and quantum nonlocality.} the particle has to be entirely in box $B_2$, at the same time as we open box $B_1$.

\item[2a.] {\it The action extends arbitrarily far\/}: the fact that the particle is entirely in box $B_2$, once we open box $B_1$, does not change with the distance between the boxes.
 
\item[2b.]  {\it The effect of that action does not decrease with the distance\/}: the effect is the creation of the particle in box $B_2$ and that effect is the same irrespective of the distance between the boxes.
 
\item[3.]  {\it This effect is individuated\/}: suppose we have a thousand boxes, each containing one particle, and that we cut each of them into two half-boxes, then send both half-boxes far apart from each other. Then, opening one half-box will affect the wave function in the other half-box (coming from the cutting in two of the same box) but not in any other half-box.
 
\item[4.] {\it That action cannot be used to transmit messages\/}: if we open box $B_1$, we learn what the wave function becomes in box $B_2$, but we cannot use that to transmit a message from the place where $B_1$ is to the one where $B_2$ is. Indeed, if one repeats the experiment many times with several boxes, one obtains that the particles are sometimes in $B_1$, sometimes in $B_2$, in an apparently random fashion, at least according to ordinary quantum mechanics. Since we have no way, by acting on one box, to choose in which of the two boxes the particle will be, it is impossible to use that experiment to send messages.
\end{itemize}
The impossibility of sending messages is sometimes taken to mean that there is nothing nonlocal going on. But nonlocality refers here to causal interactions as described (in principle) by physical theories. Messages are far more anthropocentric than that, and require that humans be able to control these interactions in order to communicate. As remarked by Maudlin, the Big Bang and earthquakes cannot be used to send messages, but they have causal effects nevertheless \cite[pp.~136--137]{Ma}.

Let us compare this kind of nonlocality with the nonlocality in Newtonian gravity. The latter also allows actions at a distance: since the gravitational force depends on the distribution of matter in the universe, changing that distribution, say by moving my body, instantaneously affects all other bodies in the universe. That action at a distance has properties 1 and 2a, but not the others. Of course, its effect decreases with the distance, because of the inverse square law, and it affects all bodies at a given distance equally (it is not individuated). On the other hand, it can in principle be used to transmit messages: if I decide to choose, at every minute, to wave my arm or not to wave it, then one can use that choice of movements to encode a sequence of zeros and ones and, assuming that the gravitational effect can be detected, one can therefore transmit a message instantaneously and arbitrarily far (but the further away one tries to transmit it, the harder the detection). Of course, all this refers to Newton's {\it theory}. There have been no experiments performed in this framework that could {\it prove} that gravitational forces really act instantaneously or at least at speeds faster than the speed of light (and, as we shall see, this is a major difference with the situation in quantum mechanics).

It is well known that Newton did not like this aspect of his own theory.\footnote{Newton thought that gravitation was mediated by particles moving at a finite speed, so that the effect of gravitation could not be instantaneous. See \cite{Ma1} for more details.} In a letter to the theologian Richard Bentley, he wrote:

\begin{quote}

 [\dots] that one body may act upon another at a distance through a vacuum without the mediation of any thing else [\dots] is to me so great an absurdity that I believe no man who has in philosophical matters any competent faculty of thinking can ever fall into it.

\begin{flushright} Isaac Newton \cite{IN, McM} \end{flushright}

\end{quote}
Not surprisingly, Einstein also firmly rejected the idea of action at a distance. In his discussions with Max Born (see Subsect.~\ref{sec6.3}), he wrote:

\begin{quote}

When a system in physics extends over the parts of space A {\it and\/} B, then that which exists in B should somehow exist independently of that which exists in A. That which really exists in B should therefore not depend on what kind of measurement is carried out in part of space A; it should also be independent of whether or not any measurement at all is carried out in space A.

\begin{flushright} Albert Einstein \cite[p. 164]{Bor} \end{flushright}

\end{quote}
In the example of the boxes, this means that opening one half-box cannot possibly influence the physical situation in the other half-box.

The Dutch physicist Hendrik Casimir underlined the fundamental problem with nonlocal actions:

\begin{quote}

If the results of experiments on free fall here in Amsterdam depended appreciably on the temperature of Mont Blanc, on the height of the Seine below Paris, and on the position of the planets, one would not get very far.
 
\begin{flushright} Hendrik Casimir \cite{Cas}, quoted in \cite{Be4} \end{flushright}

\end{quote}
If everything is connected with everything through nonlocal actions, then science becomes impossible, because, in order to test scientific theories, one always need to assume that one can isolate some systems or some variables. However, luckily, the nonlocality in quantum mechanics does not go so far as to make isolated systems impossible to realize in practice.

However, because of the problems linked with nonlocality, post-Newtonian physics has tried to eliminate property 1, while classical electromagnetism or the general theory of relativity have kept only property 2a and the negation of 4. And, due to special relativity, the combination of 1 and the negation of 4 allows in principle the sending of messages into one's own past,\footnote{At least, according to  the usual understanding of relativity. See \cite{Ma, Be4} for a more detailed discussion.} so that, if 1 holds, 4 must also hold.

One may ask whether quantum mechanics proves that there are physical effects displaying properties 1--4. The example of Einstein's boxes does not allow that conclusion, because one can consistently think that the quantum description is not complete and that the particle is always in one of the boxes.\footnote{Indeed, this is what happens in the de Broglie-Bohm theory, see  \cite{Bo1, DT, DGZ, Go-St, Bri}.} In order to prove nonlocality in the sense introduced here, i.e., a phenomenon having properties 1--4 above, we have to turn to a more sophisticated situation.

\section{A Simple Proof of Nonlocality}\label{sec4}

Let us start with an anthropomorphic thought experiment, but which is completely analogous to what happens in real experiments and could even, in principle, be realized in the anthropomorphic form presented here, as we will show at the end of this section. 

Two people, $A$ (for Alice) and $B$ (for Bob) are together in the middle of a room and go towards two different doors, located at $X$ and $Y$. At the doors, each of them is given a number, $1, 2, 3$ (let's call them ``questions", although they do not have any particular meaning) and has to say ``Yes" or ``No" (the reason why we introduce $3$ questions will become clear below). This experiment is repeated many times, with $A$ and $B$ meeting together each time in the middle of the room, and the questions and answers vary apparently at random. When $A$ and $B$ are together in the room, they can decide to follow whatever ``strategy" they want in order to answer the questions, but the statistics of their answers, although they look random, must  nevertheless satisfy two properties. We impose these properties because they are analogous to what happens in real physical experiments, but translating the experiments into an  anthropomorphic language may help us see how paradoxical our final conclusions are.

The {\it first property} is that, when the same question is asked at $X$ and $Y$, one always gets the same answer. How can that be realized? One obvious possibility is that $A$ and $B$ agree upon which answers they will give before moving toward the doors. They may decide, for example, that they will both say ``Yes'' if the question is 1, ``No'' if it is 2 and `Yes'' if it is 3. They can choose different strategies at each repetition of the experiment and choose those strategies ``at random" so that the answers will look random.

Another possibility is that, when $A$ reaches door $X$, she calls $B$ and tells him which question was asked and the answer she gave. Then, of course, $B$ can just give the same answer as $A$ if he is asked the same question and any other answer if the question is different.

But let us assume that the answers are given simultaneously, so that the second possibility is ruled out unless there exists some superluminal action at a distance between $A$ at $X$ and $B$ at $Y$. Maybe $A$ and $B$ communicate by telepathy! Of course, this is not to be taken seriously, but that is the sort of interaction that Einstein had in mind when he spoke of God using ``telepathic methods" (\cite{EinstL} quoted in \cite[p.~68]{DH}) or of  ``spooky actions at a distance" \cite[p.~158]{Bor}. 

\begin{figure}[t]
\centering 
\vspace*{25mm}
\hspace{-6.2cm}{\Large X }  {\Large  $\longleftarrow$ }  
\vspace*{-25mm}

\hspace*{20mm}    \includegraphics[keepaspectratio,height=3cm]{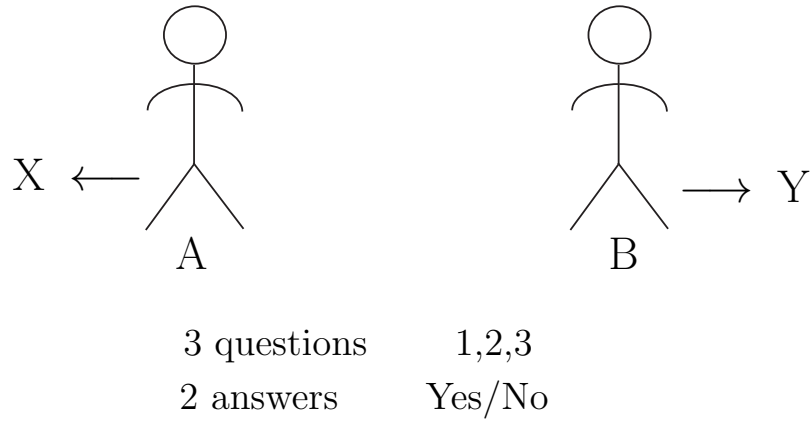}
\hspace*{40mm}    \includegraphics[keepaspectratio,height=3cm]{bonhomme.eps}

\hspace*{20mm}     {\Large A}    \hspace*{50mm} {\Large B}

\vspace*{-14mm} \hspace*{110mm} {\Large  $\longrightarrow$ }  {\Large Y} 
\vspace*{14mm}

\hspace{10mm}{\large  3 questions \qquad    1,2,3
\vspace*{2mm}

\hspace{10mm}2 answers  \qquad  Yes/No}

\caption[]{The anthropomorphic experiment}\label{4fig2}
\end{figure}

The question that the reader should ask at this point is whether there is {\it any other possibility\/}: either the answers are predetermined or (assuming simultaneity of the answers) there is a ``spooky action at a distance", namely a communication of some sort takes place between $A$ and $B$ {\it when\/} they are asked the questions. This is similar to the dilemma about the boxes: either the particle is in one of the boxes or there is some physical action between the two boxes.

Note that, to raise this dilemma, one question suffices instead of three: if the answers on both sides are always the same, then they must be predetermined if no communication is possible between the two sides. This dilemma can be called the EPR part of our argument.\footnote{However, one must mention that the original argument of  Einstein, Podolsky, and Rosen was a bit different. Einstein, Podolsky, and Rosen considered two particles moving in opposite directions and such that, because of conversation laws, one could deduce the position or the velocity of one particle by measuring the position or the velocity of the other. This is similar to the answers to the questions discussed here, except for the fact that the answers here take two values (Yes/No) instead of the continuum of values taken by the possible positions and  velocities. The version of the EPR  argument used here is due to David Bohm \cite{Bo}.}

 The reason we need three possible questions is that there is a {\it second property} of the statistics of the answers: when the two questions addressed to $A$ and $B$ are {\it different}, then the answers must be the same in only a quarter of the cases. And this property, combined with the idea that the properties are predetermined, leads to a contradiction. This is a very simple version of {\it Bell's theorem}\footnote{The argument given here is taken from \cite{DGTZ}. In the original paper by Bell  \cite{BS2}, the proof, although fundamentally the same, was more complicated.}: 

 \vspace*{5mm}

\noindent\textbf{Theorem (Bell).} We cannot have these two properties together:

\begin{itemize}
\item[1]  The answers are determined before the questions are asked and are the same on both sides (for the same questions).

\item[2] The frequency of having the same answers on both sides when the questions are different is 1/4.
\end{itemize}

\noindent\textbf{Proof.} There are three questions numbered 1, 2, and 3, and two answers Yes and No. If the answers are given in advance, there are $2^3=8$ possibilities: 
\[
\begin{array}{ccc}
1 & \qquad 2 & \qquad 3 \\
Y & \qquad Y & \qquad Y \\
Y & \qquad Y & \qquad N \\
Y & \qquad N & \qquad Y \\
Y & \qquad N & \qquad N \\
N & \qquad Y & \qquad Y \\
N & \qquad Y & \qquad N \\
N & \qquad N & \qquad Y \\
N & \qquad N & \qquad N 
\end{array}\]
In {\it each case} there are at least {\it two questions} with the same answer. Therefore,
\[
\begin{array}{l}
\mbox{Frequency (answer to 1 = answer to 2)}\\\noalign{\smallskip}
\hspace{2cm} \mbox{+ Frequency (answer to 2 = answer to 3)}\\\noalign{\smallskip}
\hspace{4cm} \mbox{+ Frequency (answer to 3 = answer to 1) $ \geq 1$}\;.
\end{array}
\]
But if
\[
\begin{array}{l}
\mbox{Frequency (answer to 1 =  answer to 2)}\\\noalign{\smallskip}
\hspace{2cm} \mbox{= Frequency (answer to  2 = answer to  3)}\\\noalign{\smallskip}
\hspace{4cm} \mbox{= Frequency (answer to  3 = answer to 1)} = 1/4\;,
\end{array}
\]
we get 
$3/4 \geq 1$, which is a contradiction.\hfill{$\blacksquare$}\vspace*{5mm}

\noindent The inequality saying that the sum of the frequencies is greater than or equal to $1$ is an example of a {\it Bell inequality}, i.e., an inequality which is a logical consequence of the assumption of pre-existing values, and which is violated by the quantum predictions.

But before drawing conclusions from this theorem, let us briefly describe how the Alice and Bob   could realize these ``impossible" statistics.
According to quantum mechanics, one can send two particles in opposite directions and with a wave function such that, if one measures the spin of those particles along the same direction on both sides (at $X$ and $Y$), one always obtains the same answer.\footnote{To be precise, the spins are anti-correlated: if one measures the spin of the particle at $X$ and one finds that it is up, the spin of the particle at $Y$ will be down. But one can decide that an up spin at $X$ means  that the answer is yes on that side,  and that an up spin at $Y$ means  that the answer is no on that side. In that way, the perfect anti-correlations are translated into getting always the same answers when one measures the spin along the same direction on both sides. One should also remark that real experiments are made with photons, where polarization plays the role played by  the spin here.} And one can choose three different directions in such a way that the
frequency of having the same answers on both sides when the directions along which the spin is measured are different is 1/4.

Then Alice and Bob can just arrange things so that such correlated particles are sent to them when they reach $X$ and $Y$. Then, they measure the spin in directions corresponding to the questions being asked to them, and give their answers according to the results of their measurements. 

It is easy to check that the nonlocality proven here does have the four properties listed in Sect.~\ref{sec3}: the results are  in principle obtained  instantaneously (or at least the time between them is much less that the one taken by the light to go from $X$ to $Y$), the correlations between the results obtained by Alice and Bob  exist no matter how far Alice and Bob are from each other and do not depend on their distance; those correlations are between two people initially together (or two particles coming from a common source) and not between one object and the rest of the Universe (as would be the case for the gravitational force) and, since neither side  can control what the results will be,\footnote{We will not give the proof of this fact, but is a well established consequence of the quantum formalism; see \cite{Eber, GRW0, Be11}, \cite[p.~139]{BH}, or \cite[Chap.~4]{Ma}.} these strange correlations cannot be used to transmit messages.

\section{Conclusions of the EPR-Bell argument}\label{sec5}

Taken by itself and forgetting about the EPR argument, Bell's result can be called a  ``no hidden variables theorem". Indeed, 
Bell showed that the mere supposition that the values of the spin pre-exist their ``measurement" (remember that those values are called hidden variables, since they are not included in the usual quantum description), combined with the perfect anti-correlation when the axes along which measurements are made are the same, and the 1/4 result for the frequencies of correlations when measurements are made along different axes, leads to a contradiction. Since the last two claims are empirical predictions of quantum mechanics that have been amply verified (in a somewhat different form), this means that these hidden variables or pre-existing values cannot exist.\footnote{Which implies that, when physicists measure the spin of a particle, they do not measure any pre-existing property of that particle. This is nicely accounted for in the de Broglie-Bohm theory, see  \cite{Bo1, DT, DGZ, Go-St, Bri}.} 
 
But Bell, of course, always presented his result {\it in combination with} the EPR argument, which shows that the  assumption of locality, combined with the perfect correlation when the directions of measurement (or questions) are the same, implies the existence of those hidden variables shown by Bell to be  impossible, because their mere existence leads to a contradiction. So, for Bell, his result, combined with the EPR argument, was not a ``no hidden variables theorem", but a nonlocality theorem, the result about the impossibility of hidden variables being only one step in a two-step argument.

To repeat, the EPR part of the argument shows that, if there are no pre-existing values, then the perfect correlations when the directions are the same imply some action at a distance. The Bell part of the argument, i.e., the theorem of the previous section, shows that the mere assumption that there are pre-existing values leads to a contradiction when one takes into account the statistics of the results when the directions are different. 

But what does this mean? It means that some action at a distance does exist in Nature, but it does not tell us what this action consists of. And we cannot answer that question without having a theory that goes beyond ordinary quantum mechanics.\footnote{The de Broglie-Bohm theory is such a theory, and non-locality is to some extent explained in that theory, see  \cite{Bo1, DT, DGZ, Go-St, Bri}.}

In ordinary quantum mechanics, what is nonlocal is the  ``collapse of the wave function": when a measurement of the spin is made at $X$ along a given direction, the wave function instantaneously collapses to a certain wave function depending on the result of the measurement but it does that {\it both at $X$ and $Y$, so nonlocally}.

Since the meaning of the wave function and its collapse is ambiguous in ordinary quantum mechanics, it is not clear that this is a real physical effect. But, as we have emphasized, if there is no physical effect whatsoever or, if one interprets the collapse of the wave function as a mere gain of information, then this means that we must have those predetermined values that lead to a contradiction.

Given the radical nature of the conclusions of the EPR--Bell argument, many attempts have been made to avoid them, i.e., to claim that the world is local after all. One such strategy is to maintain that the perfect correlations between the answers when the same questions are asked is simply a coincidence that does not need to be explained (see for example \cite{VF1, Fi1}). In the same vein, one sometimes claims that science limits itself to predictions of empirical correlations, not to explanations.

But the whole of science can be seen as an attempt to account for correlations or empirical regularities: the theory of gravitation, for example, accounts for the regularities in the motion of planets, moons, satellites, etc. The atomic theory of matter accounts for the proportions of elements in chemical reactions. The effects of medicines account for the cure of diseases, etc. To refuse to account for correlations, without giving any particular reason for doing so, is in general a very unscientific attitude. As Bell puts it:

\begin{quote}

You might shrug your shoulders and say `coincidences happen all the time', or `that's life'. Such an attitude is indeed sometimes advocated by otherwise serious people in the context of quantum philosophy. But outside that peculiar context, such an attitude would be dismissed as unscientific. The scientific attitude is that correlations cry out for explanation.

\begin{flushright} John Bell \cite[p.~152]{B}  \end{flushright}

\end{quote}
Another variant of the ``shrugging one's shoulders" argument, is to invoke a sort of ``conspiracy": for example, that each person has an answer to only one question but that, each time, and no matter how many times the experiment is repeated, that happens to be the question that is being asked to him or her. If we make that assumption, then our theorem cannot be derived (for the proof of the theorem to work, we need to assume pre-existing answers for three questions).\footnote{Another suggestion sometimes made is that the ordinary rules of probability do not apply in the EPR--Bell situation. But, as pointed out by Tumulka \cite{Tu2}, since the reasoning here relies only on frequencies of results of experiments, and since the latter obviously do satisfy the ordinary rules of probability, this attempt to ``save locality", by trying to deny the implications of the EPR--Bell argument, does not work.}

That move can be considered as an instance of what is known as  the Duhem--Quine thesis in philosophy of science: no matter what the data are, one can always save one's favorite theory (here it would be rejection of nonlocality) if one is willing to make sufficiently ad hoc assumptions. But, again, ``outside that peculiar context, such an attitude would be dismissed as unscientific". As the authors of \cite{GNTZ} observe, ``if you are performing a drug versus placebo clinical trial, then you have to select some group of patients to get the drug and some group of patients to get the placebo." But for that to work, you have to assume   ``that the method of selection is independent of whatever characteristics those patients might have that might influence how they react to the drug" \cite[Note~17]{GNTZ}. If, by accident, the people to whom the placebo is given were exactly those that are cured spontaneously, while those to whom the drug is given are so sick that the drug has little effect on them, then of course the study would be biased. And no matter how ``random" the chosen sample is, this will always remain a logical possibility. 

In any case, refusing to face a problem is not the same thing as solving it. One thing is certain: nobody has yet proposed a local explanation for those perfect correlations, and indeed nobody could do so, since Bell has shown that it is impossible.

Finally, one should emphasize that Einstein's speculations, which looked purely philosophical or even ``metaphysical" to many, have led to what is probably ``the most profound discovery of science", to use Henri Stapp's apt phrase \cite[p.~271]{Stapp}, namely the existence of nonlocal effects in the world. And the EPR and Bell papers laid the foundation for the quantum information revolution. This should be a lesson for ``pragmatists".

On the other hand, it is ironical that this result refutes Einstein's most basic assumption about Nature. As Bell said, the question raised by EPR was answered in a way that Einstein would probably ``have liked least", by showing that the ``obvious" assumption of locality made by EPR is actually false \cite[p.~11]{BS1}.

Let us now see how the physics community has, in general, misunderstood Einstein's (or EPR's) and Bell's arguments.

\section{Misunderstandings of Einstein}\label{sec6}

One rather common misconception is to think that the goal of EPR was  to show  that one could measure, say,  the spin in one direction  at $X$ and in another direction  at $Y$ for particles in the wave function described at the end of Sect.\ref{sec4}, and therefore {\it know} the values of the spin of both particles in two different directions. Since the spin along two different axes cannot be measured simultaneously, according to ordinary quantum mechanics, this would mean that the goal of EPR would be to refute that limitation in our ability to perform measurements.

But, as we explained in  Sect.\ref{sec4}, the  point of the EPR paper was not to claim that one could {\it measure} quantities that are impossible to measure simultaneously according to quantum mechanics, but rather that, if one can learn something about a physical system by making a measurement
on a distant system, then, barring actions at a distance, that ``something" must already be there before the measurement is made on the distant system.

The dilemma for Einstein was: either the quantum mechanical description of reality is complete, in which case there is something nonlocal going on in Nature, or Nature is local but the quantum mechanical description of reality is incomplete, and hidden variables have to be added to it. In his 1949 {\it Reply to criticisms\/}, Einstein did actually pose the question in the form of a dilemma, implied by the EPR argument: 

\begin{quote}

[\dots] the paradox forces us to relinquish one of the following two assertions: 

(1) the description by means of the $\psi$-function is complete,

(2) the real states of spatially separated objects are independent of each other.

\begin{flushright} Albert Einstein \cite[p.~682]{Einstein} \end{flushright}

\end{quote}

Let us now see how Bohr and  Born responded to Einstein.

\subsection{Bohr and Einstein}\label{sec6.1}

 Bohr's viewpoint not something everybody agrees on, but, from what I understand, he insisted that quantum mechanics  impose limitations to how much we can know about the microscopic world. In order to talk about that world, we need to use ``classical" concepts, meaning concepts that deal with macroscopic objects or experimental devices, and since one cannot measure mutually incompatible properties in a single experiment, one has to resort to ``complementary" views of the microscopic world: if we set up one type of experiment, we can talk about, for example, positions, and through another experiment, about momenta; or we could talk  about spins in different directions, one for each experimental setup. In each setup, we can have well-defined answers, but is is pointless to try to combine these different ``complementary" aspects of reality  into a coherent mental image. 

The root of the difference between Einstein and Bohr is that Einstein was arguing at the level of ontology, of {\it what there is\/}, insisting that a complete description of physical systems {\it must\/} go beyond the description given by ordinary quantum mechanics, if the world is local. Bohr, on the other hand, was answering systematically at the level of epistemology, of {\it what we can know\/}. Bohr was not answering Einstein, he was simply not listening to his objections. This was already the case at  the 1927 Solvay Conference, when Bohr said that  he did not  ``understand what precisely is the point which Einstein wants to [make]." ( \cite[p.~442]{BV}).

But the discussion between Bohr and Einstein centered mostly on the 1935 EPR paper. 

The ``onslaught" of EPR ``came down upon us as a bolt from the blue", wrote L\'eon Rosenfeld, who was  in Copenhagen at that time and who reported EPR's argument to Bohr. The latter then worked intensely ``week after week" on a reply.\footnote{See Rosenfeld's commentary in \cite[p.~142]{W}}. There is a  widespread belief among physicists that Bohr came up with an adequate answer to the EPR paper \cite{Bo3}. But, unlike Born, whose misunderstanding of Einstein (see Subsect.~\ref{sec6.3} below) is clearly stated, Bohr's answer is hard to understand,\footnote{As an amusing side remark, we note that, in the reprint of Bohr's paper in \cite{W}, which was the most accessible source of that paper before Internet, pages 148 and 149 are the wrong way round.} despite the fact that, according to Rosenfeld, one of his favorite aphorisms was a line from the poet Friedrich von Schiller \cite[p.~143]{W}: ``Only fullness leads to clarity, and truth dwells in the depths.''

So, what was Bohr's reply? EPR had written that:

\begin{quote}

If, without in any way disturbing a system, we can predict with certainty (i.e., with probability equal to unity) the value of a physical quantity, then there exists an element of physical reality corresponding to this physical quantity, 

\begin{flushright} Albert Einstein,  Boris Podolsky, Nathan Rosen \cite{EPR}, reprinted in \cite[pp.~138--141]{W} \end{flushright}

\end{quote}
This again means that if, by 
doing something at A, we can learn  something about the physical situation at B, 
 then what we learn must exist before we learn it, since A and B can
be far apart. Here of course, EPR assumed locality.

Bohr replied : 

\begin{quote}

[\dots] the wording of the above-mentioned criterion [\dots]
contains an ambiguity as regards the meaning of the expression ``without
in any way disturbing a system". Of course there is in a case like that
just considered no question of a mechanical disturbance of the system
under investigation during the last critical stage of the measuring procedure.
But even at this stage there is essentially the question of {\it an
influence on the very conditions which define the possible types of
predictions regarding the future behavior of the system} [\dots] their argumentation
does not justify their conclusion that quantum mechanical description
is essentially incomplete [\dots] This description may be characterized
as a rational utilization of all possibilities of unambiguous
interpretation of measurements, compatible with the finite and uncontrollable
interaction between the objects and the measuring instruments in
the field of quantum theory.

\begin{flushright} Niels Bohr \cite{Bo3}, quoted in \cite[p.~155]{B} (italics in the original) \end{flushright}

\end{quote}
Bell dissects this passage as follows:

\begin{quote}

Indeed I have very little idea what this means. I do not understand
in what sense the word `mechanical' is used, in characterizing
the disturbances which Bohr does not contemplate, as distinct from those
which he does. I do not know what the italicized passage means --- `an
influence on the very conditions [\dots]'. Could it mean just that different
experiments on the first system give different kinds of information
about the second? But this was just one of the main points of EPR, who
observed that one could learn {\it either} the position {\it or} the momentum of
the second system.\footnote{Here Bell refers to the original EPR argument discussed in  footnote 17 in Sect.~\ref{sec4}. (Note by J.B.)} And then I do not understand the final reference to
`uncontrollable interactions between measuring instruments and objects',
it seems just to ignore the essential point of EPR that in the absence
of action at a distance, only the first system could be supposed disturbed
by the first measurement and yet definite predictions become possible
for the second system. Is Bohr just rejecting the premise --- `no
action at a distance' --- rather than refuting the argument ?

\begin{flushright} John Bell  \cite[pp.~155--156]{B} (italics in the original) \end{flushright}

\end{quote}
Actually, Einstein also thought that Bohr rejected EPR's premise that ``the real situation of B could not be influenced (directly) by any measurement taken on A" (see \cite[pp.~681--682]{Einstein}). But this is by no means clearly stated and one may doubt that Bohr really understood  Einstein. 

Indeed, it is interesting to observe, as the physicist Howard Wiseman does:

\begin{quote}

When reviewing the Einstein--Bohr debates in 1949 \cite{Bohr}, Bohr concluded his
summary of his reply to EPR by quoting his defence based upon complementarity.
Astonishingly, he immediately followed this by an apology for his own `inefficiency
of expression which must have made it very difficult to appreciate the trend of the
argumentation [\dots]'! But rather than taking the opportunity to explain himself more
lucidly, he instead directed the reader to concentrate on the earlier debates with Einstein
regarding the consistency of quantum mechanics.

\begin{flushright}  Howard Wiseman \cite{Wisem}  \end{flushright}

\end{quote}
We may also note also that, in his 1949 ``Reply to criticisms" \cite{Einst1}, Einstein does not address Bohr's criticisms in any detail. Maybe this was due to the fact that he did not have  a very high opinion of Bohr's ``philosophy", at least that is the impression that one gets by reading his letters to Schrödinger, where one finds for example:

  \begin{quote}
  
  The Heisenberg-Bohr tranquilizing philosophy -- or
religion? -- is so delicately contrived that, for the time being, it provides
a gentle pillow for the true believer from which he cannot very easily be
aroused.
    
    \begin{flushright}   Albert Einstein \cite{Lett1} \end{flushright} 
   
    \end{quote}

In another letter to Schr\"odinger,  Einstein referred to Bohr as the ``Talmudic philosopher" for whom  ``reality
is a frightening creature of the naive mind" \cite{Lett2}.   Einstein also referred to Bohr as
``the mystic, who forbids, as being unscientific, an enquiry about something that exists independently of whether or not it is observed,\dots". \cite{Lett3}.

\subsection{Born and Einstein}\label{sec6.3}

Born and Einstein were lifelong friends; they both had to leave Germany, to avoid persecution, Born settling in Edinburgh, Einstein in Princeton. They exchanged a long correspondence, about physics, personal matters, and politics, which was edited by Born in 1971, long after Einstein's death in 1955 \cite{Bor}.  In that correspondence, they also discussed quantum mechanics. It is rather fascinating to read their exchange of letters and to observe the systematic misunderstanding of Einstein's point of view by Born.

For example, Einstein wrote an article in 1948 and sent it to Born (it is reproduced in the correspondence), begging him to read it as if he had just arrived ``as a visitor from Mars". Einstein hoped that the article would help Born to ``understand my principal motives far better than anything else of mine you know" \cite[p.~168]{Bor}. In that article, Einstein restated his unchanging criticisms:

\begin{quote}

If one asks what, irrespective of quantum mechanics, is characteristic
of the world of ideas of physics, one is first of all struck by
the following: the concepts of physics relate to a real outside 
world, that is, ideas are established relating to things such as 
bodies, fields, etc., 
which claim a `real existence' that is independent of the perceiving subject --- ideas which, on the other 
hand, have been brought into as secure a relationship as 
possible with the sense-data.
It is further characteristic of these physical objects that they are
thought of as arranged in a spacetime continuum. An essential aspect
of this arrangement of things in physics is that they lay claim, at a
certain time, to an existence independent of one another, provided these
objects `are situated in different parts of space'. [\dots]

The following idea characterizes the relative independence of objects
far apart in space ($A$ and $B$): external influence on $A$ has no direct
influence on $B$.

\begin{flushright} Albert Einstein \cite{Einst}, reproduced in \cite[pp.~170--171]{Bor} \end{flushright}

\end{quote}
Then Einstein repeats his usual argument, discussed above, that different choices of measurement at $A$ create (through the collapse rule) a different wave function at $B$.

Here is Born's reaction to that very same article:

\begin{quote}

The root of the difference between Einstein and me was the axiom
that events which happens in different places $A$ and $B$ are independent
of one another, in the sense that an observation on the states of affairs
at $B$ cannot teach us anything about the state of affairs at $A$.

\begin{flushright} Max Born \cite[p. 176]{Bor}  \end{flushright}

\end{quote}
As Bell says:

\begin{quote}

Misunderstanding could hardly be more complete. Einstein had no
difficulty accepting that affairs in different places could be correlated.
What he could not accept was that an intervention at one place
could {\it influence}, immediately, affairs at the other.

These references to Born
are not meant to diminish one of the towering figures of modern physics. They are meant to illustrate the difficulty of putting aside preconceptions and listening to what is actually being said. They are meant to encourage {\it you}, dear listener, to listen a little harder.

\begin{flushright} John Bell \cite[p. 144]{B} (italics in the original) \end{flushright}

\end{quote}
What Born said was that making an experiment in one place {\it teaches} us something about what is happening at another place, which is unsurprising. If, in the anthropomorphic example given in Sect.~\ref{sec4}, both people had agreed on a common strategy, one would learn what $B$ would answer to question 1, 2, or 3, by asking $A$ that same question. But, and that was Einstein's point, it would mean that the answers were predetermined and, thus, that quantum mechanics was incomplete. 

So, it seems that Born did in fact agree with Einstein that quantum mechanics is incomplete, but simply did not understand what Einstein meant by that. 


\section{Misunderstandings of Bell}\label{sec7}

As we saw, if one forgets about the EPR argument, Bell's result, can be stated as a ``no hidden variables theorem". But for Bell, his result, combined with the EPR result was not a ``no hidden variables theorem", but a nonlocality theorem, the result on the impossibility of hidden variables being only one step in a two-step argument.

 Bell was perfectly clear about this: 

\begin{quote}

Let me summarize once again the logic that leads to the impasse. The EPRB correlations\footnote{Here EPRB means EPR and Bohm, who reformulated the EPR argument in term of spins \cite{Bo}, which is the formulation used here and in most of the literature. (Note by J.B.)} are such that the result of the experiment on one side immediately foretells that on the other, whenever the analyzers happen to be parallel. If we do not accept the intervention on one side as a causal influence on the other, we seem obliged to admit that the results on both sides are determined in advance anyway, independently of the intervention on the other side, by signals from the source and by the local magnet setting. But this has implications for non-parallel settings which conflict with those of quantum mechanics. So we {\it cannot\/} dismiss intervention on one side as a  causal influence on the other.

\begin{flushright} John Bell \cite[pp. 149--150]{B} (italics in the original) \end{flushright}

\end{quote}
He was also conscious of the misunderstandings of his results:





\begin{quote}

It is important to note that to the limited degree to which {\it determinism}\footnote{Here, ``determinism" refers to the idea of pre-existing values. (Note by J.B.)}
plays a role in the EPR argument, it is not assumed but {\it  inferred}.
What is held sacred is the principle of ``local causality" --- or ``no action
at a distance". [\dots] 

It is remarkably difficult to get this point across, that determinism
is not a {\it  presupposition} of the analysis.

\begin{flushright} John Bell  \cite[p. 143]{B} (italics in the original) \end{flushright}

\end{quote}
And he added, unfortunately only in a footnote:
 
\begin{quote}

My own first paper on this subject ({\it Physics} {\bf 1}, 195 (1965))\footnote{Reference \cite{BS2}, reprinted as Chap. 2 in \cite{B}. (Note by J.B.)} starts with a summary of the EPR argument {\it from locality to} deterministic hidden variables. But the commentators have almost universally reported that it begins with deterministic hidden variables.

\begin{flushright} John Bell  \cite[p. 157, footnote 10]{B} (italics in the original) \end{flushright}

\end{quote}
An example of such a  commentator is the famous physicist Murray Gell-Mann,  Nobel Prize winner and discoverer of the  quarks, who wrote:

 \begin{quote}

 Some theoretical work of John Bell revealed
that the EPRB experimental setup
could be used to distinguish quantum mechanics from hypothetical hidden
variable theories [$\ldots$] After the
publication of Bell's work, various teams of experimental physicists
carried out the EPRB experiment. The result
was eagerly awaited, although virtually all physicists were betting on the
corrections of quantum mechanics,
which was, in fact, vindicated by the outcome.

\begin{flushright}  Murray Gell-Mann \cite[p. 172]{Ge} \end{flushright}

\end{quote}
So Gell-Mann opposes hidden variable theories to quantum mechanics, but the only  hidden variables that Bell considered were precisely those that were needed, because of the EPR argument, in order to ``save" locality. So if there is a contradiction between the existence of those hidden variables  and experiments, it is not just quantum mechanics that is vindicated, but locality that is refuted.

In one of his most famous papers, ``Bertlmann's socks and the nature of reality" \cite{BS}, Bell tried to give a pedagogical argument to explain his main idea. He gave the example of  a person (Reinhold  Bertlmann) who always wears socks of different colors.\footnote{See  \cite[p. 139]{B} for  a picture of Mr Bertlmann wearing one sock indicated ``pink" and the other ``not pink".} If we see that one sock is pink, we know automatically that the other sock is not pink (let's say it is green). That would be true even if the socks were  arbitrarily far away. So by looking at one sock, we learn something about the other sock and there is nothing surprising about that, because socks {\it do have} a color whether we look at them or not.\footnote{In case some people might worry that the notion of ``color" refers to our sensations, let us be more precise and say that the socks have the physico-chemical properties that will reflect a light of a certain wavelength (and the fact that they have those properties is  independent of our perception), which will ultimately produce our perception of color.} But what would we say if we were told that the socks have no color before we look at them? That would be surprising, of course, but the idea that quantum mechanics is complete means exactly that (if we replace the color of the socks by the values of the spin  before measurement). But then, looking at one sock would ``create'' not only the color of that sock but also the color of the other sock. And that  implies the existence of actions at a distance, if the socks are far apart. Bell emphasized that in quantum mechanics, we are in a situation similar to this nonlocal creation of colors for the socks.


However, Murray Gell-Mann made the following comment on  the Bertlmann's socks paper:

\begin{quote}

The situation is like that of Bertlmann's socks, described by John Bell in one of his papers. Bertlmann is a mathematician who always wears one pink and one green sock. If you see just one of his feet and spot a green sock, you know immediately that his other foot sports a pink sock. Yet no signal is propagated from one foot to the other. Likewise no signal passes from one photon to the other in the experiment that confirms quantum mechanics. No action at a distance takes place. (The label ``nonlocal" applied by some physicists to quantum-mechanical phenomena like the EPRB effect is thus an abuse of language. What they mean is that {\it if interpreted classically in terms of hidden variables}, the result would indicate nonlocality, but of course such a classical interpretation is wrong.)

\begin{flushright}  Murray Gell-Mann \cite[pp. 172--173]{Ge}  \end{flushright}

\end{quote}

This is not correct: it is true that, because of the random nature of the results, the experimental setup of EPRB cannot be used to send messages (or signals), as we saw in Sect.~\ref{sec3}. But nevertheless, some action at a distance does take place. And the remark about ``classical interpretation" completely misses EPR's argument showing that hidden variables are necessary if locality holds.

Of course, Gell-Mann's goal in the passage quoted here is probably to dismiss pseudo-scientific exploitations of Bell's result (such as invoking it to justify the existence of telepathy), but his defense of science is misdirected: the behavior of quantum particles is {\it not\/} like that of Bertlmann's socks (which is indeed totally unsurprising), and that was the whole point of Bell's paper.

Eugene Wigner, another Nobel Prize winner, also saw Bell's result solely as a no hidden variables result: ``In my opinion, the most convincing argument against the theory of hidden variables was presented by J.S. Bell." \cite[p. 291]{Wi}.

This is misleading, because Wigner considers only Bell's argument, which indeed shows that pre-existing spin values (or ``hidden variables") cannot exist, but forgets the EPR part of the argument, which was Bell's starting point.\footnote{See Goldstein \cite{Go1} for a further discussion of Wigner's views.}
 
To give yet another example of misunderstanding of Bell, coming from a well-known defender of the Copenhagen interpretation, consider what  Rudolf Peierls declared in an interview, referring to Aspect's experiments \cite{As} that verified the violation of Bell's inequalities:
 
\begin{quote}

If people are obstinate in opposing the accepted view they can think of many new possibilities, but there is no sensible view of hidden variables which doesn't conflict with these experimental results. That was proved by John Bell, who has great merit in establishing this. Prior to that there was a proof due to the mathematician von Neumann,\footnote{See Sect.~\ref{sec8}  for a brief explanation of von Neumann's argument. (Note by J.B.)} but he made an assumption which is not really necessary.

\begin{flushright}  Rudolf Peierls in \cite[p. 77]{ghost} \end{flushright}

\end{quote}
Basically the same mistakes were made in 1999 by one of the most famous physicists of our time, Stephen Hawking:

\begin{quote}

Einstein's view was what would now be called a hidden variables theory. Hidden variables theories might seem to be the most obvious way to incorporate the Uncertainty Principle into physics.\footnote{By this, Hawking,  means that hidden variables would be a way to maintain determinism, despite the fact that precise measurements of both position and velocities are prohibited by Heisenberg's uncertainty principle. (Note by J.B.)} They form the basis of the mental picture of the universe, held by many scientists, and almost all philosophers of science. But these hidden variable theories are wrong. The British physicist, John Bell, who died recently, devised an experimental test that would distinguish hidden variable theories. When the experiment was carried out carefully, the results were inconsistent with hidden variables.

\begin{flushright}  Stephen Hawking \cite{Haw} \end{flushright}

\end{quote}
The misconceptions about Bell continue: in an article about a Symposium held in 2000 in Vienna, in honor of the 10th anniversary of John Bell's death, 
the physicist Gerhard Gr\"ossing wrote:

\begin{quote}
 
In the 11 August 2000 issue of ``Science", D. Kleppner and R. Jackiw of MIT published a review article on ``One Hundred Years of Quantum Physics'' \cite{KJ}. They discuss Bell's inequalities and also briefly mention the possibility of ``hidden variables". However, according to the authors, the corresponding experiments had shown the following: ``Their collective data came down decisively against the possibility of hidden variables. For most scientists this resolved any doubt about the validity of quantum mechanics.''

\begin{flushright}  Gerhard Gr\"ossing \cite{GG} \end{flushright}

\end{quote}
In the February 2001 issue of {\it Scientific American\/}, Max Tegmark and John Archibald Wheeler published an article entitled {\it 100 Years of the Quantum\/} (quoted by Gr\"ossing in \cite{GG}), where one reads:
 
\begin{quote}

Could the apparent quantum randomness be replaced by some kind of unknown quantity carried out inside particles, so-called `hidden variables'? CERN theorist John Bell showed that in this case, quantities that could be measured in certain difficult experiments would inevitably disagree with standard quantum predictions. After many years, technology allowed researchers to conduct these experiments and eliminate hidden variables as a possibility. 

\begin{flushright}  Max Tegmark and John Archibald Wheeler \cite[p. 76]{TW} \end{flushright}

\end{quote}
These two last quotes clearly commit the usual mistake: they ignore the fact that Bell's result, combined with the EPR argument, refutes locality not merely the existence of ``hidden variables".

In the preface to the proceedings of the Symposium held in honor of the 10th anniversary of John Bell's death, Reinhold Bertlmann (mentioned in the ``socks" paper \cite{BS}) and the distinguished Austrian experimentalist Anton Zeilinger expressed succinctly what is probably the view of the majority of physicists about Bell's result, and which repeats the same mistake as in the previous quotes:

\begin{quote}

[\dots] while John Bell had flung open the door widely for hidden variables theories [by showing the flaws in von Neumann's proof], he immediately dealt them a major blow.  In 1964, in his celebrated paper ``On the Einstein--Podolsky--Rosen Paradox", he showed that any hidden variables theory, which obeys Einstein's requirement of locality, i.e., no influence travelling faster than the speed of light, would automatically be in conflict with quantum mechanics. [\dots] While a very tiny [experimental] loophole in principle remains for local realism, it is a very safe position to assume that quantum mechanics has definitely been shown to be the right theory.  Thus, a very deep philosophical question, namely, whether or not events observed in the quantum world can be described by an underlying deterministic theory, has been answered by experiment, thanks to the momentous achievement of John Bell.

\begin{flushright} Reinhold Bertlmann and Anton Zeilinger, preface to \cite{BZ} 
 \end{flushright}

\end{quote}

David Mermin summarized the situation described in this section in an amusing way: 

\begin{quote}

Contemporary physicists come in two varieties. 

Type 1 physicists are bothered by EPR and Bell's theorem. 

Type 2 (the majority) are not, but one has to distinguish two subvarieties.

Type 2a physicists explain why they are not bothered. Their explanations tend either to miss the point entirely (like Born's to Einstein)\footnote{See Subsect.~\ref{sec6.3}. (Note by J.B.)} or to contain physical assertions that can be shown to be false.

Type 2b are not bothered and refuse to explain why. Their position is unassailable. (There is a variant of type 2b who say that Bohr straightened out the whole business, but refuse to explain how.)

\begin{flushright}  David Mermin \cite{Me2}  \end{flushright}

\end{quote}
However, the same David Mermin also wrote: 

\begin{quote}

Bell's theorem establishes that the value assigned to an observable must depend on the complete experimental arrangement under which it is measured, even when two arrangements differ only far from the region in which the value is ascertained --- a fact that Bohm theory\footnote{See  \cite{Bo1, DT, DGZ, Go-St, Bri} for  an exposition of that theory, which is called in \cite{Bri} the de Broglie--Bohm theory. (Note by J.B.)} exemplifies, and that is now understood to be an unavoidable feature of any hidden-variables theory.

To those for whom nonlocality is anathema, Bell's Theorem finally spells the death of the hidden-variables program.  

\begin{flushright}  David Mermin \cite[p. 814]{Me4} \end{flushright}

\end{quote}
But Bell's theorem, combined with EPR, shows that nonlocality, whether we consider it anathema or not, is an unavoidable feature of the world.

Mermin adds in a footnote: ``Many people contend that Bell's theorem demonstrates nonlocality independently of a hidden-variables program, but there is no general agreement about this." \cite{Me4} It is true that there is no agreement (to put it mildly), but Mermin's own remark on the two varieties of physicists seems to explain why this is so.\footnote{What is even more surprising is that this comment comes at the end of a remarkably clear paper on the no hidden variables theorems \cite{Me4}.}

To conclude, let us highlight  some more positive reactions to Bell's result. As we mentioned at the end of Sect.~\ref{sec5}, Henry Stapp, a particle theorist at Berkeley, wrote that ``Bell's theorem is the most profound discovery of science"\cite[p. 271]{Stapp}. David Mermin mentions an unnamed ``distinguished Princeton physicist" who told him \cite{Me2}: ``Anybody who's not bothered by Bell's theorem has to have rocks in his head.'' And Feynman derived in 1982 an inequality similar to Bell's  (no reference to Bell's work is made in his paper), from which he drew the following conclusion:

\begin{quote}

That's all. That's the difficulty. That's why quantum
mechanics can't seem to be imitable by a local classical
computer.\footnote{Which means what we call locality here. (Note by J.B.)}
I've entertained myself always by squeezing 
the difficulty of quantum mechanics into a smaller and smaller
place, so as to get more and more worried about this particular item. It seems to be almost ridiculous that you
can squeeze it to a numerical question that one thing is
bigger than another.

\begin{flushright} Richard Feynman \cite[p. 485]{Fe}  \end{flushright}
\end{quote}

What Feyman calls ``one thing being
bigger than another" are inequalities of the type $1\geq 3/4$ that we used in Sect.~\ref{sec4}.

\section{Conclusions}\label{sec8}
     
We have only discussed parts of the misunderstandings that are rather widespread both among physicists and in the lay public about the history of quantum mechanics. For example, Schrödinger thought that his famous cat was a {\it reductio ad absurdum} of the idea that quantum mechanics was complete: for him, the cat was alive or dead, not  both, and since the pure quantum mechanical  description of the cat 
does not specify which is the case (the cat is in a ``superposed state"), it is manifestly incomplete or, in other words, the cat being alive or dead is another example of introducing ``hidden variables" (variables that are not part of the wave function), although in this case they are not hidden at all! Now, many people seem to think that quantum mechanics has shown that the cat is both alive and dead before anybody looks at the situation, which is the exact opposite of what Schrödinger tried to show.

Another example of misunderstandings is provided by von Neumann's no hidden variable theorem. In his 1932 book \cite{VN}, von Neumann tried to show that no hidden variables could be introduced that would complete the quantum mechanical  description. Although his proof was technically correct (and rather simple), it relied on assumptions that were quite arbitrary from a physical point of view, as shown later by Bell \cite{Be7}. But, because of the enormous (and justified) prestige of 
von Neumann as a mathematician, most physicists (including for example Max Born in  \cite[p. 109]{Born} or in the quote of Peierls cited in Sect.~\ref{sec7}) did not bother to look closely at his assumptions. Nowadays, von Neumann's theorem against the possibility of hidden variables is less often cited, but for a long time it was considered as the last nail in the coffin of the mere  idea that such variables might be introduced. 

The lesson of this sad history is that we are unlikely to understand the real issues raised by quantum mechanics, whether it be the role of the observer or nonlocality, if we continue to adhere to a history of quantum mechanics written by the victors (at least until now) and that systematically misrepresents the views of Einstein, Schrödinger, de Broglie, Bohm and Bell. Reading those authors without prejudice would already constitute a big step in the right direction.

 \end{document}